\documentclass[fleqn,10pt]{wlscirep}
\usepackage{amsmath}
\usepackage[utf8]{inputenc}
\usepackage[T1]{fontenc}
\usepackage[scaled=0.85]{beramono}
\usepackage[framemethod=TikZ]{mdframed} 
\usepackage{xcolor}
\definecolor{shadecolor}{RGB}{236,236,236}
\definecolor{matlabComment}{RGB}{60,118,61}
\definecolor{matlabbg}{RGB}{252,252,222}
\definecolor{text}{rgb}{0.6, 0.4, 0.8}    
\usepackage{textgreek}
\usepackage{graphicx}
    \usepackage[export]{adjustbox}

\newcommand{\gratingZeroPi}{\raisebox{-0.2em}{\includegraphics[scale=0.7,valign=m]{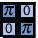}}}
\newcommand{\gratingZeroZero}{\raisebox{-0.2em}{\includegraphics[scale=0.7,valign=m]{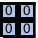}}}

\title{Quantitative phase microscopy using quadriwave lateral shearing interferometry (QLSI): principle, terminology, algorithm and grating shadow description.}

\author{Guillaume Baffou}
\affil{Institut Fresnel, CNRS, Aix Marseille Univ, Centrale Marseille, Marseille, France}

\affil{guillaume.baffou@fresnel.fr}

\affil{This is a preprint. Open access published article: \href{https://iopscience.iop.org/article/10.1088/1361-6463/abfbf9/pdf}{J. Phys. D: Appl. Phys. 54, 294002 (2021)}}

\begin{abstract}
Quadriwave lateral shearing interferometry (QLSI) is a quantitative phase imaging technique based on the use of a diffraction grating placed in front of a camera. This grating creates a wire-mesh-like image, called an interferogram, that is postprocessed to retrieve both the intensity and phase profiles of an incoming light beam. Invented in the 90s, QLSI has been used in numerous applications, e.g., laser beam characterization, lens metrology, topography measurements, adaptive optics, or gas jet metrology. More recently, the technique has been implemented on optical microscopes to characterize micro and nano-objects for bioimaging and nanophotonics applications. However, not much effort has been placed on disseminating this powerful technology so far, while it is yet a particularly simple technique. In this article, we intend to popularize this technique by describing all its facets in the framework of optical microscopy, namely the working principle, its implementation on a microscope and the theory of image formation, using simple pictures.  Also, we provide and comment an algorithm of interferogram processing, written in Matlab. Then, following the new extension of the technique for microscopy and nanophotonics applications, and the deviation from what the technique was initially invented for, we propose to revisit the description of the technique, in particular by discussing the terminology, insisting more on a grating-shadow description rather than a quadriwave process, and proposing an alternative appellation, namely "grating shadow phase microscopy" or "grating-assisted phase microscopy".
\end{abstract}
\begin{document}

\flushbottom
\maketitle
%
%
\thispagestyle{empty}



\section{Imaging the phase of light}
The wave nature of light gives a special importance to the concept of \emph{phase} in optics. In the scalar approximation, a static, monochromatic light field can be represented by a complex field $\underline{E}(\mathbf{r})=E(\mathbf{r})\exp(i\varphi(\mathbf{r}))$, where $E(\mathbf{r})\in\mathbb{R}$ is the electric field amplitude and $\varphi(\mathbf{r})\in\mathbb{R}$ is its phase, at the position $\mathbf{r}$. When probing a light field, the easily accessible physical quantity is the intensity $I(\mathbf{r})\propto|E(\mathbf{r})|^2$. But this quantity only provides a partial information. Accessing also the phase of a light field is possible but less direct, and usually requires more sophisticated techniques involving interferences, as a means to convert the phase information into measurable intensity modulations. The techniques capable of mapping the phase of a light field are coined quantitative phase imaging (QPI) techniques.

This article focuses on one particular QPI, that is named quadriwave lateral shearing interferometry (QLSI). This technique is far from being the most popular, although it gathers many advantages. This article is aimed to favor the dissemination of the technique. We first describe its principle. Then, we discuss the terminology used in QLSI introduced 20 years ago, and propose to revisit it, in order to make it simpler, more understandable and consistent, in particular for the rising applications in optical microscopy and nanophotonics. Finally, we provide and describe a Matlab code, 20-line long, suited to retrieve the intensity and phase images from a raw image camera.

\section{Principle of QLSI microscopy}

Quadriwave lateral shearing interferometry (QLSI) is a quantitative phase imaging (QPI) technique, capable of measuring both the intensity and the phase of a light beam, with high spatial resolution, and high sensitivity, in a particularly simple manner. This technology, imagined in the 90s by J. Primot,\cite{AO32_6242,JOSAA12_2679} and patented in 2000, has been mainly utilized by the community using commercialized devices sold by only one company worldwide so far (Phasics S.A.). Originally designed for precise laser beam characterization\cite{OL23_621}, the scope of applications extended in the 2000s to lens metrology, X-ray imaging,\cite{Thesis_Rizzi,Thesis_Montaux-Lambert,OE21_17340} adaptive optics and surface characterization. In 2009,\cite{OE17_13080} QLSI was proven powerful when used in an optical microscope, widening the range of QLSI applications to the study of small objects in biology\cite{OE17_13080,BJ106_1588,JBO20_126009,S14_1801910} and very recently in micro-/nanophotonics.\cite{ACSNano6_2452,PRB86_165417,ACSP4_3130,O7_243,OL38_709,PRL109_093902}

\begin{figure}[!h]
    \centering
    \includegraphics{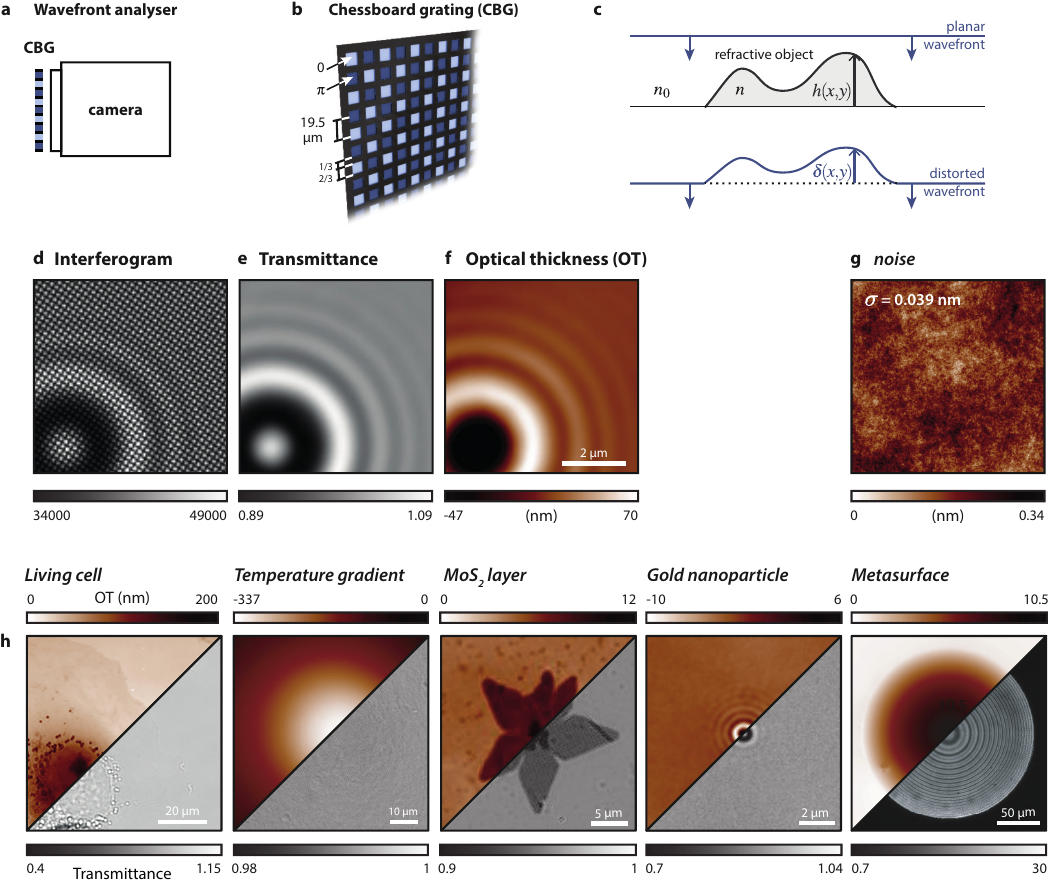}
    \caption{(a) Schematic of a QLSI wavefront analyzer, composed of a regular camera equipped with a chessboard diffraction grating. (b) Representation of a typical QLSI chessboard diffraction grating. (c) Schematic showing the wavefront distortion $\delta(x,y)$ experienced by a collimated light beam due to the presence of a metasurface. (d) Example of a raw QLSI image, called an interferogram, corresponding to a 2-\textmu m dielectric (polystyrene) bead. (d,e) Intensity and optical thickness images retrieved from the interferogram (d). (g) Image of the intrinsic noise in an OT image, with 3 s exposure time, characterized by a standard deviation of $\sigma=0.039$ nm (Zyla camera (Andor) and Sid4Element reimaging system (Phasics)). (h) Examples transmittance and OT images of micro- and nano-objects acquired by QLSI, namely a living cell,\cite{OE17_13080} a wavefront distortion created by a local micrometric induced temperature temperature in water,\cite{ACSNano6_2452} a molybdenum disulfide flake,\cite{ACSP4_3130} a 100-nm gold nanoparticle\cite{O7_243} and a metasurface.\cite{ACSP8_603}}
    \label{setup}
\end{figure}

QLSI is based on the use of a so-called \emph{wavefront analyzer} that consists of the association of two simple elements: a regular camera and a 2-dimensional (2D) diffraction grating, separated by a couple of millimeters from each other (Fig. \ref{setup}a).\cite{AO39_5715} The grating affects both the phase and the intensity of light: Opaque horizontal and vertical lines are blocking the light, and defining transmitting square areas imprinting phase shifts of $0$ and $\pi$ on the transmitted light, according to a chessboard pattern (Fig. \ref{setup}b).\cite{AO39_5715} Upon illumination, the diffraction grating creates an image, called an interferogram (Fig. \ref{setup}d), on the camera sensor that can be processed to retrieve both the intensity (Fig. \ref{setup}e) and the wavefront profile $W(x,y)$ (Fig. \ref{setup}f) of a light beam . When the wavefront distortion $W$ originates from an imaged object (Fig. \ref{setup}c), the mapping of $W$ can be used to optically characterize the object. In this case, instead of wavefront profile, one rather speaks about the optical path difference (OPD), or equivalently the optical thickness (OT) of the object $\delta\ell=W$, defined as
\begin{equation}
\delta(x,y)=(n-n_0)h(x,y),
\end{equation}
where $n$ is the refractive index of the object and $n_0$ the refractive index of the surrounding medium (Fig. \ref{setup}c). When the object is not uniform, this more general expression applies:
\begin{equation}
\delta(x,y)=\int_0^{h(x,y)}(n(x,y,z)-n_0)\mathrm{d}z.
\end{equation}
One also sometimes refers to the phase of the light $\varphi(x,y)$, which can be calculated from $W$ provided the illumination wavelength $\lambda_0$ is known:
\begin{equation}
\varphi=\frac{2\pi}{\lambda_0}W\label{ellphi}
\end{equation}
Thus, calling QLSI a \emph{phase} imaging technique is somehow inconsistent, as it primarily measures a wavefront profile. Strictly speaking, it is more a wavefront sensing technique than a QPI. Retrieving the phase requires a preknowledge (the wavelength), and the phase may not be accurately defined in the case of a broadband illumination. All these physical quantities, phase, OPD, OT and wavefront profile, are equivalent, interchangeable, and all used in the literature.\\

The advantages of QLSI over other existing QPI techniques are many-fold. (i) An interesting advantage is the achromaticity. Counterintuitively, although based on the use of a grating, not only the technique can be used with broadband illumination, but also the knowledge of the illumination wavelengths is not necessary to calculate the intensity and wavefront images from the interferogram (the reason is explained later on). This feature makes QLSI particularly robust. (ii) The noise standard deviation in OPD measurements is typically 0.6 \AA$\cdot$Hz$^{-1/2}$, i.e., $\sim10^{-4} \lambda_0\cdot$Hz$^{-1/2}$  corresponding to around 1 mrad$\cdot$Hz$^{-1/2}$ of phase delay in the visible range. The identified sources of noise in the reconstructed phase images are inhomogeneities, shot noise and non-linearities of the camera.\cite{AO32_6242} (iii) It provides an evaluation of the noise from the measurement itself (see Refs.\cite{AO32_6242,OE21_17340} and discussion in Section \ref{sect:demodulation})  (iv) The spatial (i.e., lateral) resolution reaches the diffraction limit, unlike other popular wavefront sensors such as Shack-Hartmann. (v) It is easy to use, as it just consists in using a camera-like device. No modification of the microscope is required. (vi) QLSI benefits from the high sensitivity of interferometric methods but do not suffer from their usual drawbacks: it neither requires a reference beam, nor a complex alignment that might be sensitive to external perturbations. The relative positioning of the grating with respect to the camera is done once and for all, and is not sensitive to, e.g., temperature variation, mechanical drift or air flow.

\begin{figure}[!h]
    \centering
    \includegraphics{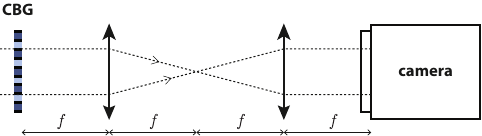}
    \caption{Relay-lens configuration to remotely position the chessboard grating (CBG) in front of the camera sensor. $f$ is the common focal length of the two lenses.}
    \label{relaylens}
\end{figure}

Note that the implementation of a diffraction grating around 1 mm from the sensor of a camera is not always possible, due to the frequent presence of a vacuum, or atmosphere-controlled, chamber in front on the sensor. This caution is used to avoid the presence of water and crystallization upon cooling down the camera sensor. In that case, the diffraction grating has to re-imaged using a relay lens, consisting of a $4-f$ optical system (Fig. \ref{relaylens}). The use of a reimaging system introduces chromaticity, because it makes the effective distance between the grating and the sensor wavelength-dependent. This dependence has to be characterized to achieve quantitative measurements (see discussion on the $\alpha$ factor in Equation \ref{GammaEquation}).

Albeit conceived in the 90s, the idea of plugging a QLSI device on a microscope was only introduced in 2009,\cite{OE17_13080} in the context of using QLSI as a quantitative phase microscopy technique for biological cell imaging (Fig. \ref{setup}h). Following this pioneering work, QLSI has been applied in bioimaging for tissue retardance imaging,\cite{OC422_17} imaging of various organelles,\cite{JBO17_076004} dry biomass measurements,\cite{JBO20_126009,JCB211_765} microtubule visualization,\cite{BJ106_1588} neurite growth quantification,\cite{A146_1361} nanolocalization,\cite{NC6_7764} and superresolution microscopy.\cite{NM15_449,FP7_68}

In parallel, QLSI has also been used in the field of nanophotonics, to image the temperature distribution around metal nanoparticles under illumination,\cite{ACSNano6_2452,S14_1801910} to optically characterize 2D-materials,\cite{ACSP4_3130} single nanoparticles,\cite{PRL109_093902,O7_243}, non-linear coherent Raman signals on microparticles,\cite{OL38_709,PRL109_093902} and metasurfaces \cite{ACSP8_603} (Fig. \ref{setup}h).

All the above-mentioned studies, albeit diverse, have been conducted by only a limited number of research groups. The current cost of commercial QLSI devices and the long-standing myth that QLSI is sophisticated presumably explains the reluctance of the community to buy the technology, or to set it up. Moreover, the complexity of the name "quadriwave lateral shearing interferometry" may be reluctant and help perpetuate the myth of an unaccessible technique. The aim of this article is to favor the dissemination of the technique, make it more accessible, and contribute to provide QLSI with the attention it deserves, in particular by breaking this myth.

\section{Interferogram formation}
All the magic of QLSI happens between the grating and the camera, over a couple of millimeters. What is happening in there can be understood in two ways. The first picture deals with the interference between four diffraction orders of the grating, hence the name of the technique. The second picture, less popular albeit simpler, refers to a shadowing effect of the grating, very similar to the working principle of a Shack-Hartman wavefront sensor. Both of them are as important. Depending on the community or the application under consideration, one picture could be more appropriate than the other. In this section, focus in put on the understanding of the working principle of QLSI, using simple pictures.

\subsection{The Grating Shadow (GS) picture}

\begin{figure*}[h]
    \centering
    \includegraphics[scale=1]{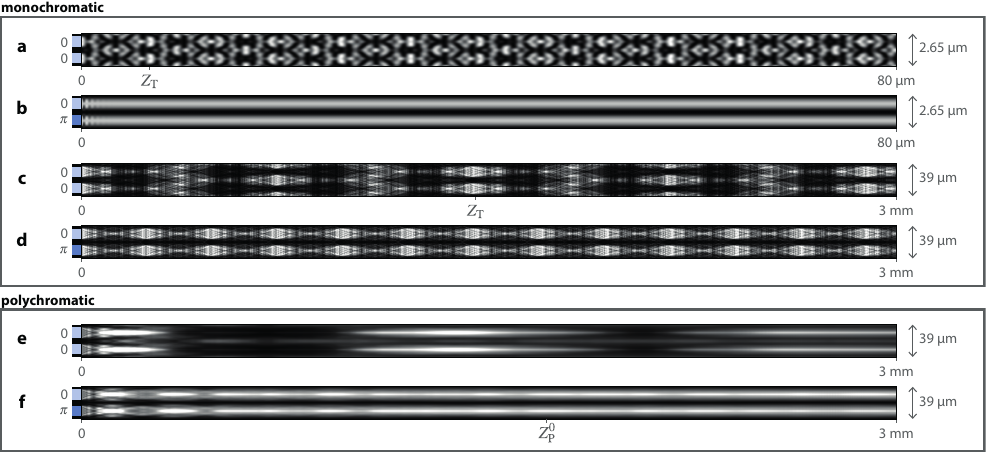}
    \caption{Modelling of the light propagation (from left to right) after various diffraction grating geometry. (a) Grating period: $\Lambda=2.65/2$ \textmu m, $\lambda_0=550$ nm, no chessboard phase pattern. (b) $\Lambda=2.65$ \textmu m, $\lambda_0=550$ nm, $0-\pi$ chessboard phase pattern. (c) $\Lambda=39/2$ \textmu m, $\lambda_0=550$ nm, no chessboard phase. (d) $\Lambda=39$ \textmu m, $\lambda=550$ nm, $0-\pi$ chessboard phase pattern. (e) $\Lambda=39/2$ \textmu m, $\lambda_0=[450,650]$ nm, no chessboard phase. (f) $\Lambda=39$ \textmu m, $\lambda_0=[450,650]$ nm, $0-\pi$ chessboard phase pattern. }
    \label{ChessBoardGratingPatterns2}
\end{figure*}

The $0-\pi$ chessboard-pattern of the grating, depicted in Fig. \ref{setup}b, is the key feature to make QLSI work. This phase shift arrangement makes the diffraction grating \emph{diffraction-less},\cite{AO39_5715} as strange as it may seem. Figure \ref{ChessBoardGratingPatterns2} explains this behavior by showing calculations of the light propagation between the grating and the camera, for different grating properties, from simple to complex. Calculations have been performed for a single unit cell of the grating (such as \gratingZeroPi), with periodic conditions. Figure \ref{ChessBoardGratingPatterns2}a starts with a grating without $0-\pi$ alternation (unit cell: \gratingZeroZero). The transmitted light features complex and contrasted interferences, as expected for a grating. A periodicity of the pattern in the $z$ direction can be observed, a property coming from the Talbot effect, stating that the grating is re-imaged at periodic distances separated by
\begin{equation}
Z_\mathrm{T}=2\Lambda^2/\lambda_0.
\end{equation}
where $\Lambda$ is the grating period. Interestingly, when the $0-\pi$ phase shift alternation \raisebox{-0.2em}{\includegraphics[scale=0.7,valign=m]{0piUnitCell-eps-converted-to.pdf}} is restored (Fig. \ref{ChessBoardGratingPatterns2}b), then the interferences fully disappear, and the transmitted light becomes invariant by translation along $(Oz)$, propagating like a normal shadow behind an opaque object. This peculiar behavior comes from the cancellation of the zero diffraction order by forward destructive interferences, so that only the four 1st orders of the transmitted light remain. The transmitted light pattern is thus equivalent to the interferences observed after a Fresnel biprism, in 2 dimensions. These two introductory cases (Figs. \ref{ChessBoardGratingPatterns2}a,b) do not exactly match commonly used QLSI gratings, since the grating periodicity in these calculations was very small ($\Lambda=1.325$ \textmu m). QLSI wavefront sensors rather feature a periodicity corresponding to 6 to 8 times the periodicity of the camera sensor (i.e., typically $6\times6.5=39$ \textmu m) in order to resolve the interferogram modulation. For $\Lambda=39$ \textmu m, the shadowing effect depicted previously remains if a $0-\pi$ chessboard pattern is applied (Fig. \ref{ChessBoardGratingPatterns2}d), but the beam exhibits some inner structuration. Interestingly, these structurations disappear once a polychromatic beam is used. Each wavelength creates its own structuration, and they all cancel each other out after a given distance that reads\cite{Thesis_Rizzi,Thesis_Montaux-Lambert}
\begin{equation}
Z_\mathrm{P}=\sqrt{-\frac{\ln(V_0)}{2\pi^2}}\frac{\Lambda^2}{\Delta\lambda_0},
\end{equation}
where $\Delta\lambda_0$ is the spectral width of the illumination and $V_0$ the tolerated blurring of the interference pattern ($0<V_0<1$). For an arbitrary value of $V_0=1/e$ (corresponding to 30\% of contrast), one gets the simple expression:
\begin{equation}
Z_\mathrm{P}^0=-\frac{1}{\sqrt{2}\,\pi}\frac{\Lambda^2}{\Delta\lambda_0}
\end{equation}
This effect, depicted by Primot et al. in 2000,\cite{OC180_199,OE21_17340,Thesis_Rizzi,Thesis_Montaux-Lambert} was coined the panchromatic Talbot effect, and $Z_\mathrm{P}$ the panchromatic distance. This shadowing effect and invariance by $z$-translation makes the grating-camera distance not so critical. The separation distance can be continuously varied over a few millimeters. It is not necessary to place the grating above the panchromatic distance $Z_p^0$. Below $Z_p^0$, the high frequencies observed in Fig.\ref{ChessBoardGratingPatterns2}f can't be resolved by the camera anyway, in most cases. Increasing the distance between the camera and the grating has the advantage to make the technique more sensitive. However, it also makes it less suited to image phase gradients that exhibit high spatial frequencies.

\begin{figure}[!h]
    \centering
    \includegraphics{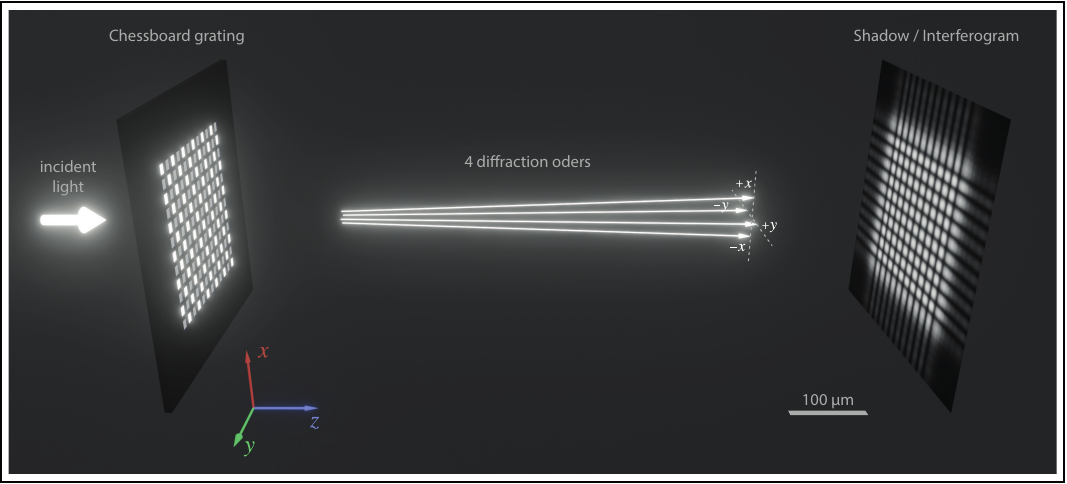}
    \caption{Overview of the working principle of QLSI. The size of the diffraction grating has been reduced to $6\times6$ unit cells for the sake of clarity. The incident optical wavefront is decomposed in 4 wavefronts propagating along the four 1st diffraction orders of the grating. These diffraction orders interfere to create a fringe pattern, that perfectly mimics the geometrical shadow of the grating. In this 3D scheme, the camera-grating distance was 1 mm. The displayed interferogram is a numerical calculation considering a camera-grating distance of 1 mm, a wavelength range of 450-650 nm and a grating period of 39 \textmu m.}
    \label{quadriwave}
\end{figure}

Figure \ref{quadriwave} gives an overview of the general working principle of QLSI. The size of the diffraction grating has been reduced to $6\times6=36$ unit cells for the sake of clarity. In practice, it is rather composed of more than $300\times300$ unit cells, to cover the size of the camera sensor. The camera (here positioned at 1 mm from the grating) collects a shadow-like image of the grating, composed of periodic dots arising from the interferences between four diffraction orders, two along the $x$ direction and two others along the $y$ direction, the zero order being cancelled by destructive interferences created by the $0-\pi$ chessboard pattern. Also, the $2/3$ ratio of the squares depicted in Fig. \ref{setup}b optimizes the dimming of higher orders of diffraction.\cite{AO44_1559} This splitting in four directions of propagation creates 4 images on the camera sensor, only slightly shifted with each other typically by an angle of $\xi=0.5^\circ$, following Bragg's law, $2\Lambda\sin\xi=\lambda_0$, where $\Lambda=39$ \textmu m is the grating period.

\begin{figure*}[!h]
    \centering
    \includegraphics{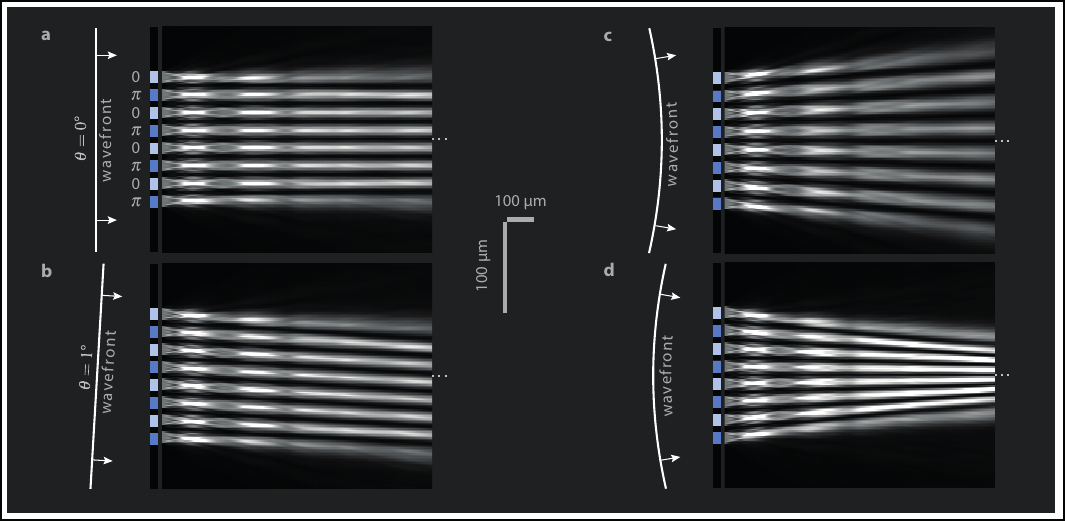}
    \caption{The QLSI principle understood as a shadowing effect. (a) Calculation of the transmitted light intensity after a $0-\pi$ chessboard grating, for a planar incoming light wavefront at normal incidence. For the sake of simplicity, the square grating contains $4\times4$ unit cells. A diffraction-less shadowing effect is observed. (b) Same calculation when the wavefront is tilted by an angle of $1^\circ$, producing a lateral shift of the shadow according to this same angle. (c) Simulation of transmitted light when the wavefront is convexe, producing a spreading of the shadow. (d) Simulation of transmitted light when the wavefront is concave, producing a shrinking of the shadow. }
    \label{gratingDeviation}
\end{figure*}

Let us focus now on the effect of a wavefront deviation from a planar shape on the interferogram, which is what primarily matters. Figure \ref{gratingDeviation} presents numerical simulations of light propagation between the grating and the camera sensor for flat, tilted, and curved optical wavefronts. For the sake of simplicity, numerical simulations were conducted for a grating made of $4\times4$ unit cells. A titled wavefront shifts the interferogram, while a curved wavefront dilates or shrinks it (Fig. \ref{gratingDeviation}c,d). Of course, with a real grating composed of 10000s of unit cells, these deformations occur locally, enabling the mapping of complex 2D wavefront profiles, with high spatial resolution. Thus, the information regarding the phase gradient is contained in the displacement of the dots of the interferogram, in the exact same manner as in a Shack-Hartmann wavefront sensor. The advantage of QLSI is that the density of dots is much higher, optimized, does not require the fabrication of an array of microlenses, and enables the processing of a wavefront image with as many pixels as the camera sensor.\footnote{Note that some algorithms reduce the number of pixels in the phase image by a factor of 9 or 16 to accelerate image processing.}

\begin{figure*}[!h]
    \centering
    \includegraphics{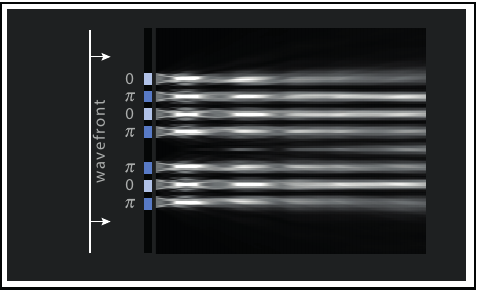}
    \caption{Calculation of the transmitted light intensity after a $0-\pi$ chessboard grating, for a planar incoming light wavefront at normal incidence, where one hole of the grating is blocked. A self-healing of the beam is evidenced, showing the limitation of the shadowing picture.}
    \label{gratingDarkPixel}
\end{figure*}

The shadowing picture depicted above provides a simple explanation of the working principle of the technique, but has limitations. Figure \ref{gratingDarkPixel} shows what happens when one hole of the grating is blocked. As expected, no light is transmitted outside this hole, following the shadow picture, but after a few 100s of microns, the light beam is recovered. This effect is usually called "beam self-healing" in optics. Thus, this simple shadow picture should be used with caution. It can be used to popularize the principle making it understandable by the layman, simply explain why the technique is achromatic (see next section), etc, but not to properly investigate the underlying physics and advance science. The light propagation after the grating remains a diffraction process and to carry out fundamental research in grating-assisted phase microscopy, the multiwave picture has to be considered.

\subsection{Influence of the wavelength}
Interestingly, the shifts of the dots represented in Fig. \ref{gratingDeviation} do not depend on the wavelength, only on the wavefront gradient, as it mimics a simple shadowing effects. This is the reason why QLSI is achromatic, although it is based on a diffraction grating. The knowledge of the wavelength is not even necessary to reconstruct the wavefront image from the interferogram. Only the grating-camera distance matters.

However, it does not mean that the wavelength has no effect on the interferogram quality, and that a given QLSI wavefront sensor can be used for any wavelength without problem. To imprint the $\pi$ phase shifts in the chessboard grating, the transparent substrate is locally etched to remove a thickness $h$ such that $\pi=\frac{2\pi}{\lambda_0}(n_\mathrm{s}-1)h$. For glass, one has $n_\mathrm{s}\approx1.5$, which gives $h=\lambda_0$. Thus, a diffraction grating is made for a specific wavelength a priori. If used for another wavelength, the phase shifts deviate from $\pi$. The principle depicted in Fig. \ref{gratingDeviation} still applies, but the contrast of the interferogram will be reduced. In other words, the measures are still quantitative, but the signal to noise ratio is poorer and the sensitivity reduced. However, this limitation is not dramatic. Typical QLSI wavefront sensors suited for visible wavelengths can be used in the 450-750 nm range without much problem. In the Fourier space discussed in the next section, an unadapted wavelength produces additional diffraction spots, that can still be removed numerically, if need be.

\section{Discussion of the terminology}

\subsection{QLSI \emph{vs} GS}

How come a simple grating in front of a camera can be coined quadriwave lateral shearing interferometry? This question is often raised by the layman and the answer is not trivial, as detailed in the previous section. Our experience is that this terminology may even be reluctant and give a negative prejudice regarding the complexity of the method, especially for the new communities starting using this QPI technique (namely optical (bio)microscopy and nanophotonics): the complexity of the name "quadriwave lateral shearing interferometry" does not reflect the simplicity of the technique, experimentally speaking. The reason of the name QLSI is historical: The technique resembles a previous QPI technique named lateral shearing interferometry (LSI), based on the interferences of two light beams, tilted with each other using a prism. But LSI does not involve a grating and is very different experimentally speaking from QLSI. We believe that, at some point, one should detach from historical considerations and focus on scientific considerations, to reach a better description. Otherwise, descriptions are made at the expense of the simplicity and clarity. 

This chessboard grating is the central (and even the only) part of the QLSI technique. The grating is what makes this technique unique, different from all the other phase imaging and wavefront sensing techniques. For this reason, the use of the word "grating" within the name of the technique would be justified and natural. "Quadriwave lateral shearing interferometry" neither tells what the technique is (a grating in front of a camera), nor what it does (wavefront sensing or a phase imaging).

For these reasons, names such as "Grating shadow (GS) phase microscopy", "Grating shadow (GS) wavefront sensing" or "Grating-assisted phase microscopy" would certainly ease the dissemination of the technique, its reference, its explanation, and would contribute to break the myth of a sophisticated and out-of-reach technique for the layman. Changing the name of a 20 year-old scientific field, device, technique or concept is difficult, albeit not impossible. Many examples exist. It usually takes time and gives rise to some reluctance from the initial community, following the famous Planck's principle.



\subsection{Grating \emph{vs} mask}
Until here, in this article, we have been using the name "grating" to define the diffraction element, which seems natural. Yet, the community usually prefers the name "modified Hartmann mask".\cite{OE29_1239,OA39_5715} We also propose here to revisit this appellation to favor a better understanding. First, the grating has little to do with a Hartmann mask. Originally, a Hartman mask is not even a grating. A Hartman mask consists of an opaque screen pierced by several holes (originally three, sometimes more), to help adjust the focus of a telescope. Second, the main aim of the chessboard grating is not to \emph{mask} part of the incoming beam. On the contrary, it is to engineer the transmitted light. Thus, it makes more sense to call it a grating rather than a mask. Again, the reason of the "mask" appellation is more historical than scientific, making it difficult to understand.

Born \& Wolf, in their seminal book,\cite{book_BornWolf} defined a grating as "any arrangement which imposes on an incident wave a periodic variation of amplitude or phase, or both", which is exactly what the chessboard grating does.  Interestingly, they also noted that the common analysis of a one-dimensional grating "may easily be extended to two- and three-dimensional periodic arrangements of diffracting bodies", but they stated that, unlike 1D or 3D gratings, "2D gratings (called cross-gratings) found no practical applications". It seems time has changed. Note that a 1D diffraction grating can also be used in a similar optical configuration as a QPI. This technique is called 'diffraction phase microscopy' \cite{Bhaduri6_57}, but does not benefit from the interest of measuring both wavefront gradients to achieve better wavefront reconstruction.

Interestingly, the use of a grating is not even necessary. It certainly contributes to optimize the interferogram quality, but non-periodic, and even random, optical elements have been shown efficient. A particularly simple and cheap approach consists in replacing the grating with a thin diffuser, creating a speckle pattern on the camera sensor, instead of the well-defined, periodic arrangement of dots. Then, the principle just consists in monitoring the distortion of the speckle pattern to retrieve the wavefront profile. This technique has been recently pioneered for optical \emph{microscopy} developments,\cite{OL42_5117} but was proposed long ago for more general applications.\cite{OE29_1239}

\subsection{Talbot \emph{vs} Bessel}

The translational invariance of the transmitted light after a chessboard grating (provided a polychromatic light is used, see Fig. \ref{ChessBoardGratingPatterns2}f) has been coined the "panchromatic Talbot effect". However, a Talbot effect is rather a phenomenon of image replication at periodic distances, not an invariance by translation. What happen between the grating and the sensor rather belongs to the family of propagation-invariant optical waves, the most famous members of this family being the Bessel beams. The nature of the optical wave could thus be referred as a Bessel-like wave, produced when a polychromatic light is used. Once again, the reason of the reference to a Talbot effect is historical, because it has been discovered upon studying the Talbot effect.



\section{Image retrieval algorithm}

\begin{figure}[!h]
    \centering
    \includegraphics{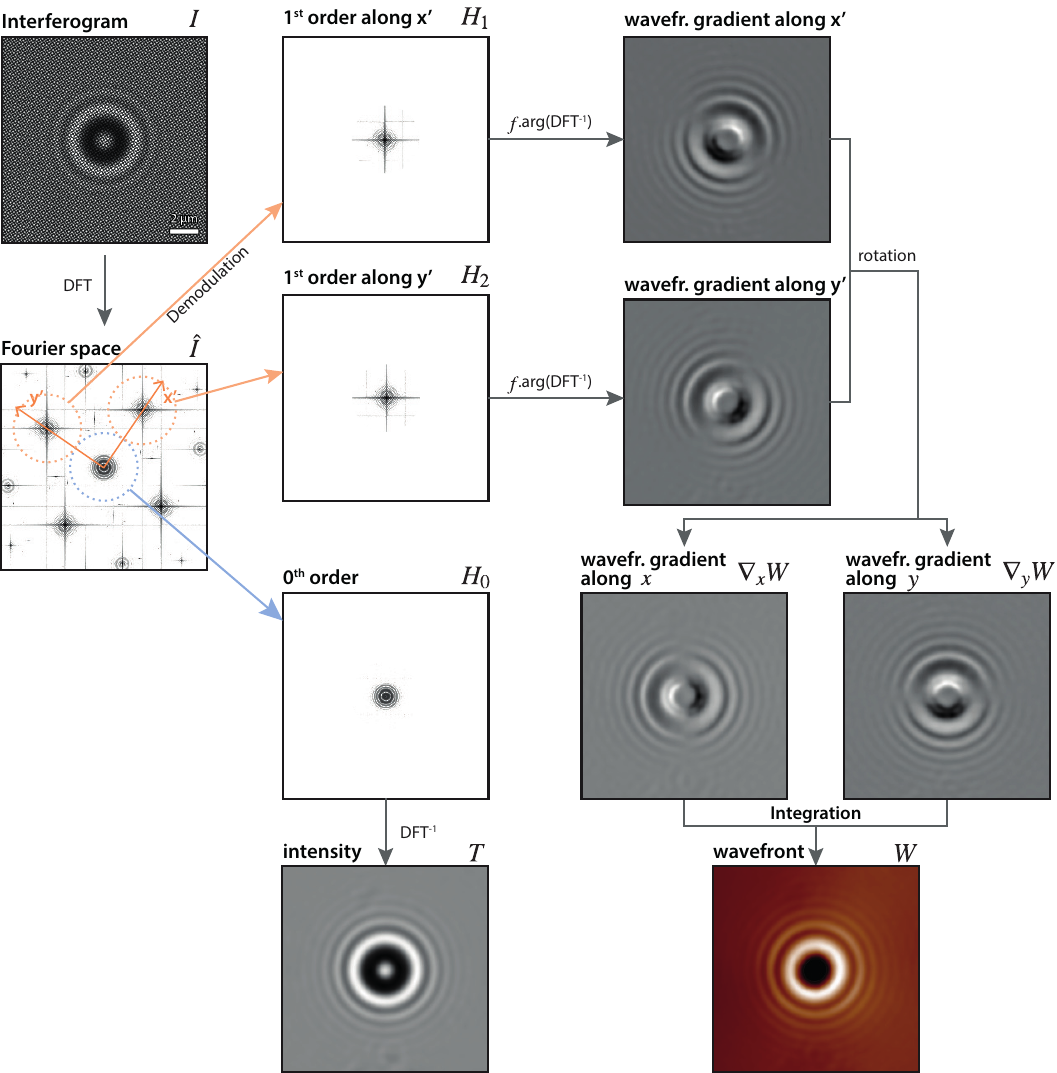}
    \caption{Summary schematic of the full intensity and wavefront retrieval algorithm, illustrated with the image of a 2-\textmu m dielectic bead.}
    \label{QLSIprocedure}
\end{figure}

The section is aimed to show that numerically retrieving the phase (or equivalently the wavefront profile) of light in GS microscopy is simple. A description of this procedure can be found in Refs. \cite{Thesis_Rizzi,Thesis_Montaux-Lambert} (in French). The image displayed in Figure \ref{setup}d is called the interferogram. It is the raw data acquired by the camera sensor, and resembles the intensity image (Figure \ref{setup}e), on top of which the shadow of the grating looks printed. When the incoming wavefront is perfectly planar, the fringes of the grating's shadow are perfectly parallel. When the incoming wavefront is non uniform, an imperceptible distortion of these fringes appears, which creates deviations from a perfect spatial periodicity that can be extracted by a demodulation algorithm involving Fourier analysis. This is the basic principle of the retrieval algorithm that we shall explain in this section. The overall numerical procedure is schematized in Figure \ref{QLSIprocedure}, with the example of the image of a 2-\textmu m dielectric (polystyrene) bead (see Fig. \ref{setup}d). Let us break it down to explain the different steps, and illustrate them using Matlab codes.\\


\subsection{Fourier transform}
Let $I$ be the interferogram image.  The first step consists in calculating the Fourier transform of the interferogram:
\begin{equation}
\hat I=\mathcal{F}[I]
\end{equation}

\begin{mdframed}
	[font=\tt,
	backgroundcolor=matlabbg,
	frametitle=Matlab code,
	frametitlefont=\tt,
	frametitlerule=true,
	frametitlebackgroundcolor=shadecolor]       
> \hspace{1em} I=imread(\textcolor{text}{'interferogram.tif'});\\
> \hspace{1em} Ihat=fftshift(fft2(I));
\end{mdframed}
\begin{figure}[h]
    \centering
    \includegraphics[scale=1]{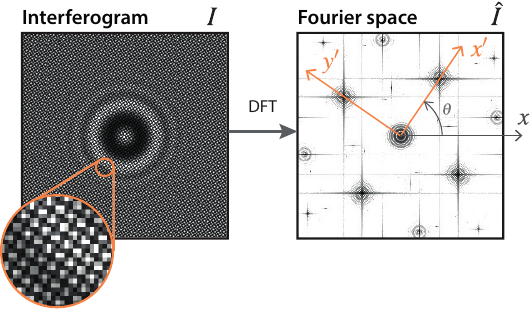}
    \caption{(left) Raw interferogram $I$ acquired by the camera and (right) representation of its Fourier transform $\hat I$.}
    \label{algo_FourierTransform}
\end{figure}
Figure \ref{algo_FourierTransform} displays the discrete Fourier transform (DFT) of the interferogram. In addition to the central spot, four main peripheral Fourier spots are visible. They correspond to the fringes of the interferogram, and contain the information on the wavefront profile of the light beam. The positions of these spots define two directions $(x')$ and $(y')$. These directions are tilted by an angle $\theta$ compared with the original $(Oxy)$ frame  because the grating is tilted by this angle. All the diffraction spots have a 4-branch star shape, with branches along the x and y directions. Without tilt, the branches of the zero order could overlap 1st diffraction orders, which would create artefacts, hence the tilt. In this image, the distance between the central spot and a peripheral spot is $1/3$ of the size of the image. This is because the fringe periodicity ($\Lambda/2$, half the grating period) equals three times the pixel size. This configuration optimizes spatial resolution of the processed phase and intensity images.

\subsection{Demodulation\label{sect:demodulation}}
The second step consists in isolating the 1st order diffraction spots. This is where the information related to the wavefront profile is contained. There are 4 spots, i.e., 4 options a priori. However, diametrically opposed spots are redundant. They contain exactly the same information since they are exact complex conjugate. This symmetry in the Fourier space comes from the fact that the original image $I$ is real. Thus, only two spots can be considered, without loss of information, any of them provided they are $90^\circ$ apart.

\begin{figure}[h]
    \centering
    \includegraphics[scale=1]{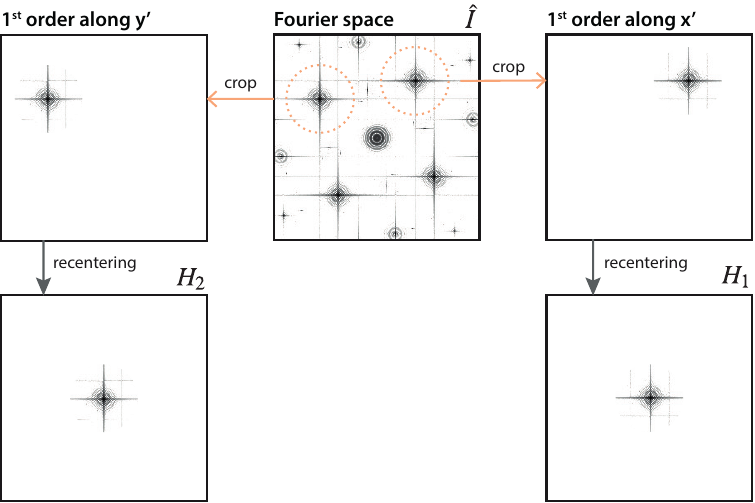}
    \caption{Schematic of the demodulation procedure in the Fourier space.}
    \label{algo_demodulation}
\end{figure}

For both spots, the process is the same. The spot is cropped by a disc of radius $R_\mathrm{c}$ (or a square). This disc should be small enough, to avoid gathering information from neighboring spots, but not too small to avoid too much removal of high spatial frequencies of the image, and a blurring of the final wavefront image. $R$ typically equals one sixth of the image. If the image is not square, then the cropped area is an ellipse. Then, the cropped spot is centered in the Fourier space (Figure \ref{algo_demodulation}).

\begin{mdframed}
	[backgroundcolor=matlabbg,font=\tt]
> \hspace{1em} \textcolor{matlabComment}{\textbf{\%\% Demodulation of a diffraction spot}}\\
> \hspace{1em} \textcolor{matlabComment}{\% Crop of the diffraction spot}\\
> \hspace{3em} [Ny,Nx] = size(Ihat);\\
> \hspace{3em} [xx,yy] = meshgrid(1:Nx,1:Ny);\\
> \hspace{3em} dx = 127; dy = 210; ~~\textcolor{matlabComment}{\% position of the first order spot to be cropped}\\
> \hspace{3em} R2C = (xx-Nx/2-1-dx).$\hat{~}$2/Rcx$\hat{~}$2 + (yy - Ny/2-1-dy).$\hat{~}$2/Rcy$\hat{~}$2;\\
> \hspace{3em} circle = (R2C < 1); ~~\textcolor{matlabComment}{\% mask circle}\\
> \hspace{3em} Ihatc = Ihat.*circle;\\
> \hspace{1em} \textcolor{matlabComment}{\% Recentering of the cropped spot}\\
> \hspace{3em} H1=circshift(Ihatc,[-dy -dx]);
\end{mdframed}

This procedure is to be repeated for the second spot, leading to two images that we name $H_1$ and $H_2$, corresponding respectively to the spots along $x'$ and $y'$.

This crop procedure in the Fourier space yields a reduction of the spatial frequencies in the image, by a factor of around 3 (half the grating pitch $\Lambda/2$ equals three times the camera pixel size). Consequently, the spatial resolution of the intensity and phase images in QLSI is reduced by a factor of 3 compared with the image that the camera would measure without grating. Thus, to reach the diffraction limit in phase and intensity in QLSI, the image has to be oversampled compared with the diffraction limit by at least of factor of 3. Note that some commercial QLSI cameras use of factor of 4, which decreases the spatial resolution.

Note that not only diffraction spots along the $x'$ and $y'$ are observed, but also along $x'+y'$ and $x'-y'$. These other diffraction spots could be used to retrieve the wavefront profile as well, and could even be used for this reason to get more signal and improve the signal to noise ratio. Interestingly, this redundancy could also be used to estimate the error of the measurement, as explained in Refs. \cite{AO32_6242,OE21_17340}.

\subsection{Inverse Fourier transform}
The images $H_1$ and $H_2$ shall be inverse-Fourier transformed. Back in the $(x,y)$ space, the fringes are now gone, since the spot was centered in the Fourier space. However, The recovered image has no reason to be real anymore, since the crop and translation in the Fourier space cancelled the Hermitian property of the Fourier image. The values of the recovered images are thus complex and, interestingly, the \emph{argument} of these complex values is proportional to the local wavefront gradient $\nabla W(i,j)$. The proportionality constant depends on the periodicity $\Lambda$ of the chessboard grating, and on its distance $d$ to the sensor:\cite{Thesis_Rizzi}
\begin{eqnarray}
\nabla_{x'}W&=&\frac{\Lambda}{4\pi d}\arg\left(\mathcal{F}^{-1}[H_1]\right)\equiv\alpha\arg\left(\mathcal{F}^{-1}[H_1]\right),\nonumber\\
\nabla_{y'}W&=&\frac{\Lambda}{4\pi d}\arg\left(\mathcal{F}^{-1}[H_2]\right)\equiv\alpha\arg\left(\mathcal{F}^{-1}[H_2]\right).\label{GammaEquation}
\end{eqnarray}
In practice, the prefactor $\alpha$ is not calculated from the knowledge of $\Lambda$ and $d$. $\Lambda$ is precisely known, but $d$ cannot really be measured precisely as the space between the camera and the grating is not easily accessible. The factor $\alpha$ is rather determined experimentally, using a reference sample of known optical thickness (e.g., grooves on a glass sample characterized by AFM).

When using a relay-lens system (see Figure \ref{relaylens}), the distance $d$ between the image of the grating and the camera usually depends on the wavelength due to unavoidable achromaticity of the relay lens. For this reason, the calibration factor $\alpha(\lambda)$ is wavelength dependent and has to be determined over the full wavelength range of interest.

Before integrating the vectorial gradient to get the wavefront profile, one needs to rotate it by an angle $-\theta$ to retrieve gradients over the axes $(x)$ and $(y)$:
\begin{equation}
	\begin{bmatrix}
		\nabla_{x}W \\ \nabla_{y}W
	\end{bmatrix}=
	\begin{bmatrix}
		\cos(\theta)& -\sin(\theta) \\ \sin(\theta)&\cos(\theta)
	\end{bmatrix}
	\cdot
	\begin{bmatrix}
		\nabla_{x'}W \\ \nabla_{y'}W
	\end{bmatrix}
\end{equation}

This is a requirement as any vectorial gradient integration algorithm (see below) assumes gradients over x and y, i.e., along the horizontal and vertical line of the image matrix.

\begin{figure}[h]
    \centering
    \includegraphics[scale=1]{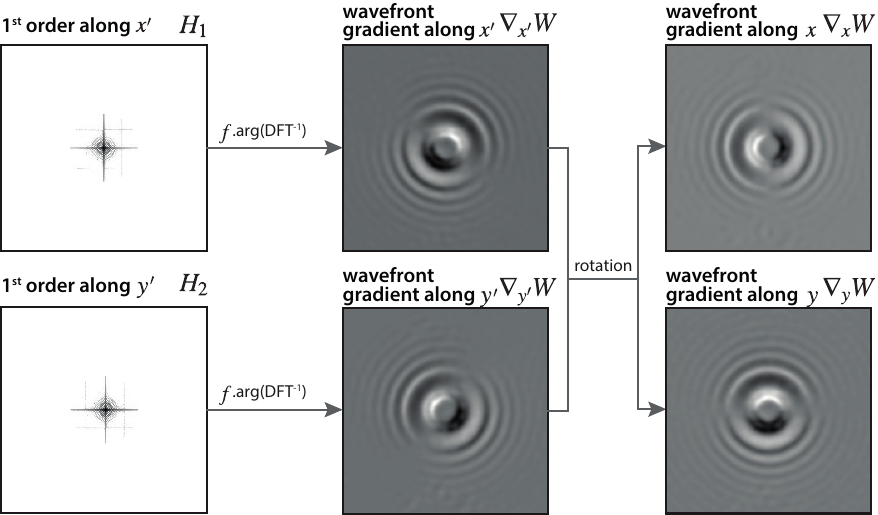}
    \caption{Schematic of the retrieval of the wavefront gradients from the Fourier images $H_1$ and $H_2$.}
    \label{algo_InverseFourier}
\end{figure}

\begin{mdframed}
	[font=\tt,
	backgroundcolor=matlabbg]       
> \hspace{1em} \textcolor{matlabComment}{\textbf{\%\% Inverse Fourier transform}}\\
> \hspace{1em} Ix = ifft2(ifftshift(H1));\\
> \hspace{1em} Iy = ifft2(ifftshift(H2));\\
> \hspace{1em} alpha = -0.00318;\\
> \hspace{1em} \textcolor{matlabComment}{\textbf{\%\% Wavefront gradient calculation}}\\
> \hspace{1em} DW1 = alpha*angle(Ix);\\
> \hspace{1em} DW2 = alpha*angle(Iy);\\
> \hspace{1em} \textcolor{matlabComment}{\textbf{\%\% Wavefront gradient rotation}}\\
> \hspace{1em} DWx = cos(theta)*DW1-sin(theta)*DW2;\\
> \hspace{1em} DWy = sin(theta)*DW1+cos(theta)*DW2;
\end{mdframed}

This part of the algorithm is depicted in Figure \ref{algo_InverseFourier}.

\subsection{Wavefront gradient integration}
\begin{figure}[!h]
    \centering
    \includegraphics[scale=1]{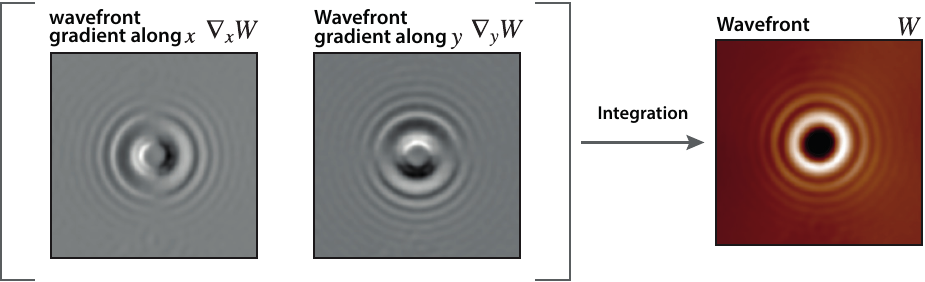}
    \caption{Schematic of the integration of the wavefront gradients to retrieve the wavefront image.}
    \label{algo_GradientIntegration}
\end{figure}

The final step consists in a very fundamental mathematical task: retrieving a 2D scalar field from its vectorial gradient (Fig. \ref{algo_GradientIntegration}). Several algorithms are freely availabe, some of them written in Matlab. 
For instance, the Matlab package from John D'Errico does a good job.\cite{ErricoMatlabPackage}

\begin{mdframed}[font=\tt,backgroundcolor=matlabbg]    
> \hspace{1em} \textcolor{matlabComment}{\textbf{\%\% Wavefront gradient integration using the John D'Errico algorithm}}\\   
> \hspace{1em} DWx=DWx-mean(DWx(:));\\
> \hspace{1em} DWy=DWy-mean(DWy(:));\\
> \hspace{1em} W = p*intgrad2(DWx,DWy);
\end{mdframed}

\noindent where $p$ is the camera pixel size. The removal of the mean values of $\nabla_{x}W$ and $\nabla_{y}W$ prior to gradient integration are meant to suppress the rather aleatory offset that yields an unphysical ramp on the integrated image.

Although this wavefront retrieval postprocessing works perfectly fine in 99\% of cases, it may not yield proper reconstruction for three reasons: (i) The wavelength is far from the nominal wavelength the grating has been designed for. Indeed, the $\pi$ phase shift of the grating are produced by local substrate etching, the depth of which equals to wavelength to ensure a phase shift of $\pi$. The use of a different wavelength from the nominal wavelength does not create any bias, since QLSI is achromatic. However, it will decrease the signal to noise ratio until a point where the wavefront image will no longer be nicely reconstructed. Typically, a single grating can work fine from 480 to 750 nm. (ii) If the object under study is optically too thick and sharp, and creates a phase step in the image that exceeds $\pi$, then the wavefront gradients may not be properly reconstructed. It comes from the fact that Eqs. \eqref{GammaEquation} involve the argument of a complex number, which is defined modulo $2\pi$. However, it does not mean phase differences within the image cannot exceed $\pi$. For instance, the technique can image living cells, a few microns (optically) thick without any issue. (iii) If the wavefront distribution does not derive from a gradient, then the algorithm naturally yields problems. In principle, when imaging refractive objects, the OPD is simply related to the thickness of the object, and the wavefront profile derives from a gradient. However, when using unusual samples (complicated phase plates or metasurfaces with vortices for instance), or unusual samples illuminations (speckle, vortex beam, etc), then the phase profile may feature singularities. Interestingly, M. Guillon et al. recently fixed this issue by developing an algorithm capable of dealing with the presence of phase vortices.\cite{arXiv:2101.07114}

\subsection{Intensity image retrieval}
\begin{figure}[h]
    \centering
    \includegraphics[scale=1]{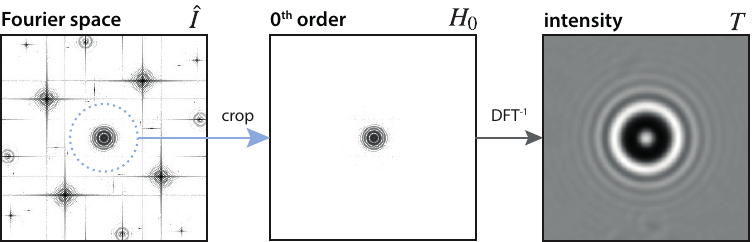}
    \caption{Schematic of the intensity image retrieval alrogithm.}
    \label{algo_Tretrieval}
\end{figure}

Not only the phase (or equivalently the wavefront profile) can be retrieved from the interferogram, the intensity image can also be calculated using a similar procedure. This time, the central spot (zero order) has to be cropped in the Fourier space, by the same disc of radius $R$ (Fig. \ref{algo_Tretrieval}).

\begin{mdframed}
	[backgroundcolor=matlabbg,font=\tt]
> \hspace{1em} \textcolor{matlabComment}{\% Crop of the diffraction spot}\\
> \hspace{1em} R2C = (xx-Nx/2-1).$\hat{~}$2/Rcx$\hat{~}$2 + (yy - Ny/2-1).$\hat{~}$2/Rcy$\hat{~}$2;\\
> \hspace{1em} circle = (R2C < 1); ~~\textcolor{matlabComment}{\% mask circle}\\
> \hspace{1em} H0 = Ihat.*circle;\\
> \hspace{1em} T = ifft2(ifftshift(H0));
\end{mdframed}

\subsection{Dealing with the reference image}
The way it is presented above, the procedure is slightly simplified. Used this way, the procedure would give a wavefront image tarnished by a lot of imperfections, coming from microscope aberrations, pieces of dust on the optics, light beam imperfection. All these static wavefront imperfections can be, and must be, removed by acquiring another interferogram image, usually called the reference. In microscopy experiments, the reference is usually acquired on a blank field of view, free from any object of interest. If no blank area is present in the sample (like in the case of cultured cells in confluence for instance), then the reference image can be acquired upon moving the sample holder rapidly and randomly during the exposure time. The whole algorithm detailed above must be also applied to the reference to retrieve the intensity reference $T_\mathrm{ref}$ and, more importantly, the wavefront reference $W_\mathrm{ref}$. $T_\mathrm{ref}$ can be used to calculate the transmittance image $t=T/T_\mathrm{ref}$. $W_\mathrm{ref}$ must be calculated and subtracted to the OPD image retrieved from the above-detailed algorithm in order to extract a proper wavefront profile $W$.

In practice $W$ and $W_\mathrm{ref}$ are not subtracted at the end of the code. The subtraction can be done at the moment when the wavefront gradients along $(Ox')$ and $(Oy')$ are calculated:

\begin{mdframed}
	[font=\tt,
	backgroundcolor=matlabbg]       
> \hspace{1em} \textcolor{matlabComment}{\textbf{\%\% Phase gradient calculation}}\\
> \hspace{1em} DW1 = factor*angle(Ix.*conj(Ix\_ref));\\
> \hspace{1em} DW2 = factor*angle(Iy.*conj(Iy\_ref));\\
\end{mdframed}

\section{Summary}

In light of recent applications in optical microscopy, in particular for biomiaging and nanophotonics, we propose here to revisit the field quadriwave lateral shearing interferometry, with simple working principle descriptions and some new terminology. In particular, we propose to use grating-shadow phase microscopy as a well-suited name of the technique, which should favor its dissemination, ease its description, and contribute to make it more popular and accessible for these fields of applications. We explained the working principle of GS phase microscopy and highlight the main advances when implemented on an optical microscope, in particular in bioimaging and nanophotonics for nanoparticle, 2D material and metasurface characterization. We go more into detail about the working principle by describing what occurs between the grating and the camera, the only two elements involved in GS phase microscopy. In particular, we explain that two complementary visions can be considered: a 4-image interference picture, and a grating-shadow picture. Finally, we detail the image retrieval algorithm, and provide related Matlab codes.

\section*{Acknowledgments}
The author wishes to thank Daniel Andr\'en, Rodrigo Guti\'errez-Cuevas and Miguel A. Alonso for helpful discussions, and Ljiljana Durdevic for providing the U2OS cell image of Figure 1h. This work has received funding from the European Research Council (ERC) under the European Union's Horizon 2020 Research and Innovation Programme (grant agreement no. 772725, project HiPhore).


\end{document}